\documentstyle[sprocl]{article}

\def\CQG{{\it Class. Quantum Gravity} }

\def\PL{{\it Phys. Lett.} }
\def\PR{{\it Phys. Rev.} }
\def\PRL{{\it Phys. Rev. Lett.} }

\def\UFN{{\it Usp. Fiz. Nauk} }

\def\SPU{{\it Soviet. Phys. Usp.} }

\def\be{\begin{equation}}
\def\ee{\end{equation}}
\def\bea{\begin{eqnarray}}
\def\eea{\end{eqnarray}}
\def\be{\beta}

\def\vev#1{\langle {#1}\rangle}

\def\frac#1#2{{\textstyle{{#1}\over {#2}}}}

\def\lsim{\mathrel{\rlap{\lower4pt\hbox{\hskip1pt$\sim$}}
    \raise1pt\hbox{$<$}}}
\def\gsim{\mathrel{\rlap{\lower4pt\hbox{\hskip1pt$\sim$}}
    \raise1pt\hbox{$>$}}}
\def\sqr#1#2{{\vcenter{\vbox{\hrule height.#2pt
	 \hbox{\vrule width.#2pt height#1pt \kern#1pt
	 \vrule width.#2pt}
	 \hrule height.#2pt}}}}

\begin{document}

\title{SEED MAGNETIC FIELDS FROM THE BREAKING OF LORENTZ INVARIANCE}

\author{ O. BERTOLAMI \footnote{Talk presented 
at the Meeting on CPT and Lorentz Symmetry, Bloomington, USA, 
November 1998.} }

\address{Departamento de F\'\i sica, Instituto Superior T\'ecnico,\\
Av. Rovisco Pais, Lisbon, Portugal}

\author{D.F. MOTA}

\address{University of Cambridge, DAMTP, Silver Street, CB3 9EW, U.K.}

\maketitle\abstracts{
Spontaneous breaking of Lorentz invariance may take place in 
string theories, possibly endowing the photon with a mass. 
This leads to the breaking of the conformal symmetry of the
electromagnetic action allowing for 
the generation within inflationary scenarios of  magnetic 
fields over $Mpc$ scales. We show that the 
generated fields are consistent with amplification by  
the galactic dynamo processes and can be  
as large as to explain the observed galactic magnetic 
fields through the collapse of protogalactic clouds.}

\section{Introduction}

Magnetic field of nearby 
galaxies coherent over $Mpc$ scales are estimated to be of order 
$B \sim 10^{-6}$ G.
The most plausible explanation for these 
fields involves some sort of dynamo effect. 
Indeed, if one assumes that a 
galactic dynamo has operated during about 10 Gyr then a seed magnetic field 
could be amplified by a factor of $e^{30}$ and the observed 
galactic magnetic fields at present may have had its origin in a 
seed magnetic field of about $B \sim 10^{-19}G$. On the other hand,
the galactic magnetic fields can emerge directly from the 
compression of a primordial magnetic field, in the collapse of 
protogalactic clouds. In this case, it is required a seed magnetic field of 
$B \sim 10^{-9}G$ over a scale $\lambda \sim Mpc$, the comoving size 
of a region which condenses to form a galaxy. Since the Universe through most 
of its history has behaved as a good conductor it 
implies that the evolution of any primeval cosmic magnetic field will 
conserve magnetic flux. Thus, the ratio denoted 
by $r$, of the energy 
density of a magnetic field $\rho_B = {B^2 \over 8\pi}$ relative to the 
energy density of the cosmic microwave background radiation 
$\rho_\gamma = {\pi^2 \over 15} T^4$ remains essentially constant and 
provides a invariant measure of 
magnetic field strength.
It then follows that pregalactic magnetic fields of about 
$r \approx 10^{-34}$ are required if one invokes dynamo 
amplification processes, and 
$r\approx 10^{-8}$ if one assumes only the collapse of protogalactic clouds.

In what follows we shall describe a mechanism to generate primordial
magnetic fields that is based on a putative violation of the
Lorentz invariance in string field theory solutions and that 
relies on inflation
for the amplification of quantum fluctuations of the electromagnetic field
\cite{paper}.
Invoking a period of inflation to explain the creation of 
seed magnetic fields is a quite attractive suggestion as inflation 
provides the means of generating large-scale phenomena from 
microphysics that operates on subhorizon scales \cite{turner}. 
Indeed, inflation, 
through de Sitter-space-produced quantum fluctuations, generates 
excitations of the electromagnetic field allowing for an increase 
of the magnetic flux before the Universe is filled with a highly conducting 
plasma. Furthermore, in this amplification process, long-wavelength modes 
for which $\lambda \geq\ H^{-1}$, are enhanced. 

However, it is not possible to produce the required seed magnetic fields 
from a conformally invariant theory as is 
the usual U(1) gauge theory. The reason being that, in a 
conformally invariant theory, the magnetic field decreases as 
$a^{-2}$, where $a$ is the scale factor, and during inflation, 
the total energy density in the Universe is 
constant, so the magnetic field energy density is strongly suppressed, 
yielding $r = 10^{-104}\lambda_{Mpc}^{-4}$.

In the context of string field theory, there exists solutions where 
conformal invariance may be broken, 
actually due to the possibility of spontaneous breaking of the Lorentz 
invariance \cite{kostelecky}. 
This possibility arises explicitly from solutions 
of string field theory of the open bosonic string, as interactions are cubic 
in the string field and these give origin in the static field theory potential
to cubic interaction terms of the type $SSS$, $STT$ and $TTT$, where 
$S$ and $T$ denote scalar and tensor fields. Lorentz invariance 
may then be broken as can be seen, for instance, from   
static potential involving the tachyon, $\vev{\varphi}$, 
and a generic vector field \cite{kostelecky}.
The vacuum of this model is unstable and  
gives rise to a mass-squared term for the vector field that is proportional 
to $\vev{\varphi}$. If $\vev{\varphi}$ is negative, 
then the Lorentz symmetry itself is spontaneously broken as the vector field
can acquire a non-vanishing vacuum expectation values. 
This mechanism can give rise to vacuum expectation values to tensor 
fields inducing for the fields that do not acquire vacuum expectation values,
such as the photon \cite{kostelecky1}, mass-squared terms proportional
to $\vev{T}$. Hence, one should expect from this 
mechanism terms for the 
photon of the form $\vev{T}A_{\mu}A^{\mu}$, 
$\vev{T_{\mu \nu}}A^{\mu}A^{\nu}$ and so on. Naturally, these terms break
explicitly the conformal invariance of the electromagnetic action. 

Observational constraints on the breaking of the Lorentz invariance 
arising from the measurements of the quadrupole splitting time 
dependence of nuclear Zeeman levels along Earth's orbit, 
have been performed over the years, the most recent one indicating that  
$\delta < 3 \times 10^{-21}$ \cite{chupp}.
Bounds on the violation of momentum conservation and  
existence of a preferred reference frame can be also 
extracted from limits on the
parametrized post-Newtonian parameter $\alpha_{3}$ obtained
from the period of millisecond pulsars, namely
$|\alpha_{3}| < 2.2 \times 10^{-20}$ \cite{bd} 
implying the Lorentz symmetry is unbroken  
up to this level. These limits indicate that if the Lorentz invariance is
broken then its violation is suppressed by powers of energy over the string
scale. Similar conclusions can be drawn for possible violations of the
CPT symmetry \cite{potting}.

In order to relate the theoretical possibility of spontaneous breaking of
Lorentz invariance to the observational limits discussed above 
we parametrize
the vacuum expectation values of the Lorentz tensors in the following way:

\begin{equation}
\label{parametrization}
<T> = m_{L}^2 \left({E \over M_{S}}\right)^{2l}~,
\end{equation}
where $m_L$ is a light mass scale when compared to string typical
energy scale, $M_{S}$, where we assume that $M_{S} \approx M_{P}$, 
$M_P$ being the Planck mass; 
$E$ is the temperature of the Universe in a given period and $2l$ is 
a positive integer. We shall further replace the temperature of
the Universe by the inverse of the scale factor, given that expansion of 
the Universe is adiabatic. Parametrization 
(\ref{parametrization}) is similar to the 
one used in previous work \cite{potting,bertolami1,bertolami2}. 

\section{Generation of Seed Magnetic Fields}

We consider spatially flat Friedmann-Robertson-Walker 
cosmologies with the metric given in the conformal time, $\eta$, the 
corresponding scale factor being, $a(\eta)$, and 
the stress tensor of a perfect fluid. 
The Hubble constant is written as 
$H_0 = 100 \ h_0~km~s^{-1}~Mpc^{-1}$ and the present Hubble radius is 
$R_0 = 10^{26}\ h_0 ^{-1} \ m$, where $0.4\leq h_0 \leq 1$. 
We shall assume the Universe has gone through a period of 
expontential inflation at a scale $M_{GUT}$ and whose
associated energy density is given by $\rho_I \equiv M_{GUT}^{4}$. 
Hence, from the Friedmann equation, 
$H_{dS} = ({8\pi \over 3})^{1/2}~{M_{GUT}^2 \over M_P}$.

From our discussion on the breaking of Lorentz 
invariance we consider for simplicity only the term, 
$\vev{T}A_{\mu}A^{\mu}$, from which follows the 
Lagrangian density for the photon:

\begin{equation}
\label{lagrangian}
{\cal L} = - \frac{1}{4} F_{\mu \nu} F^{\mu \nu} + M_L^2 a^{-2l} A_\mu A^\mu ,
\end{equation}
where $M_L^2 \equiv {m_L^2 \over M_p^{2l}}$.
One can readily obtain the wave equation for the magnetic field:

\begin{equation}
\label{motionB}
{1 \over a^2} {\partial^{2} \over \partial \eta^2} a^2 \vec B - \nabla ^2 
\vec B + {n \over \eta^2} \vec B = 0~.
\end{equation}

The corresponding equation for the Fourier components of 
$ \vec B$ is given by:

\begin{equation}
\label{motionF}
\ddot {\vec F_k} + k^2 \vec F_k + \frac{n}{\eta ^2} \vec F_k = 0~,
\end{equation}
where the dots denote derivatives according to the conformal time and 
$\vec F_k(\eta) \equiv a^2 \int{ d^3x e^{i \vec k . \vec x} \vec 
B( \vec x, \eta)}$,
$\vec F_k$ being a measure of the magnetic flux associated with the
comoving scale $ \lambda \sim k^{-1} $. The energy 
density of the magnetic field is given by 
$\rho_B(k) \propto {|\vec F_k|^2 \over a^4}$.

For modes well outside the horizon, $a \lambda >> H^{-1}$ or $|k \eta| << 1$, 
solutions of Eq. (\ref{motionF}) are given in terms of the conformal time 
\cite{turner}:

\begin{equation}
\label{Feta}
| \vec F_k| \propto \eta ^{m_{ \pm}}
\end{equation}
where $m_{ \pm} = \frac{1}{2} \left[1 \pm \sqrt{1-4n}\right]$.

By requiring that $n$ is not a growing function of conformal time, 
it follows that $n$ has to be either a constant 
or that $2l$ is negative, which is excluded by our assumption 
(\ref{parametrization}). Hence for different phases of evolution 
of the Universe:

\noindent
(I) Inflationary de Sitter (dS) phase, where $a( \eta) \propto - 
{1 \over \eta H_{dS}}$, it follows that $l = 0$ and

\begin{equation}
\label{sittern}
n = - \frac{M_{dS}^2}{H_{dS}^2}~~,
\end{equation}
where we refer $M_L$ by its index in the relevant phase of 
evolution of the Universe.

\noindent
(II) Phase of Reheating (RH) and Matter Domination (MD), where $a( \eta) 
\propto \frac {1}{4} H_0^2 R_0^3 \eta ^2$, yields from the 
condition $n$ is a constant that $2l = 3$ and

\begin{equation}
\label{MDRHn}
n = - \frac{4M_{MD}^2}{H_0^2R_0^3}~~. 
\end{equation}

\noindent
(III) Phase of Radiation Domination (RD), where $a( \eta) \propto 
H_0 R_0^2 \eta $, from which follows that $l = 2$ and

\begin{equation}
\label{RDn}
n = - \frac{M_{RD}^2}{H_0^2R_0^4}~~. 
\end{equation}
It is clear that in this case last case $n \ll 1$.

Assuming the Universe has gone through a period
of inflation at scale $M_{GUT}$ and that fluctuations of the electromagnetic 
field have come out from the horizon when the Universe had gone through about
$55$ $e$-foldings of inflation, yields in terms of $r$ \cite{turner}:

\begin{equation}
\label{ratio}
r \approx (7 \times 10^{25})^{-2(p+2)} \times (\frac{M_{GUT}}{M_{P}})
^{4(q-p)/3} \times (\frac{T_{RH}}{M_{P}})^{2(2q-p)/3} 
\times (\frac{T_{*}}{M_{P}})^{- 8q/3} \times 
\lambda_{Mpc}^{-2(p+2)},
\end{equation}
where $T_{*}$ is the temperature at which plasma effects become dominant 
and can be estimated from the reheating process \cite{turner}
$T_{*} = min \{(T_{RH}M_{GUT})^{ \frac{1}{2}};(T_{RH}^2 M_{p})^{ 
\frac{1}{3}}\}$. For the reheating temperature 
we assume either a
poor or a quite efficient reheating, $T_{RH} = \{ 10^9~GeV ; M_{GUT} \}$. 
Finally, $p \equiv m_{-dS} = {1 \over 2} \left[1 - \sqrt{1+
\left({2M_{dS} \over H_{dS}}\right)^2}\right]$ and $q \equiv m_{+ RH} 
= {1 \over 2} \Biggl[1 + \sqrt{1+ 16 {M_L^2 \over H_0^2 R_0^3}}\Biggr]$ 
are the fastest growing solutions 
for $\vec F_k$ in the de Sitter and reheating phases, respectively. 

In order to obtain numerical estimates for $r$ we have to compute $M_L$. 
At the de Sitter phase we have that 
$M_{dS} = m_{dS}$. As we have seen $m_L$ is a 
light energy scale when compared to $M_P$ and 
$M_{GUT}$, and hence we introduce a parameter, $\chi$, 
so that $m_{dS} = \chi M_{GUT}$ and $\chi \ll 1$. 

At the matter domination phase, we have to impose that the mass term 
$M_{MD} = m_{MD} ({T_{\gamma} \over M_{p}})^{l}$, $T_{\gamma}$ 
being the temperature of the cosmic background radiation at about 
the recombination time, is consistent with the 
present-day limits of the photon mass, 
$m_{\gamma} \leq 3 \times 10^{-36}~GeV$ \cite{chibisov}. Thus, 
at the matter domination phase, we have to satisfy the condition, 
$M_{MD} \leq m_{\gamma}$ which implies that 
$m_{MD} ({T_{\gamma} \over M_{P}})^{3/2} \leq 3 \times 10^{-36}~GeV$, 
following that $m_{MD} \leq 7.8 \times 10^{4}~GeV$.
A more stringent bound on $m_{MD}$ could be obtained from the limit
$m_{\gamma} \leq 1.7 \times 10^{-42}~h_{0}~GeV$ arising from the
absence of rotation in the polarization of light of distant galaxies 
due to Faraday effect \cite{carrol}.

We present in the following table our estimates for the ratio $r$
for $M_{GUT} =10^{16}~GeV$. One can see that we obtain values that 
are in the range $10^{-35} < r < 10^{-9}$, where a poor reheating and 
the lower values for $\chi$ tend to render $r$ 
too low even for an amplification via dynamo processes.
Estimates for different values of $M_{GUT}$ can be found in Ref. [1]. 

\begin{center}
{\bf Table }
\end{center}

\begin{table}[!h]
\centering
\begin{tabular}{|l|l|l|l|l|l|} \hline\hline
${ \em \chi }$ & ${ \em p}$ & ${ \em q}$ & ${ \em T_{RH}(GeV)}$ 
& ${ \em T_{*}}(GeV)$ & ${ \em \log~r}$  \\ \hline\hline $5 \times 10^{-3}$ 
& $-1.67$ & $1$ & $10^{9}$ & $2.3 \times 10^{12}$ 
& $-35$  \\ \hline $5 \times 10^{-3}$ & $-1.67$ & $1$ & $10^{16}$ & $10^{16}$ 
& $-27$  \\ \hline
$6 \times 10^{-3}$ & $-2.08$ & $1$ & $10^{9}$ & $2.3 \times 10^{12}$ 
& $-18$  \\ \hline
$6 \times 10^{-3}$ & $-1.08$ & $1$ & $10^{16}$ & $10^{16}$ & $-9$  \\ \hline

\end{tabular}
\caption{Values of $r = \frac{\rho_B}{\rho_\gamma}$ at 
$1~Mpc$ for $M_{GUT} = 10^{16}~GeV$}
\end{table}

\section{Summary}

We have shown that the strength of the magnetic field 
produced by considering the spontaneous breaking of the Lorentz
symmetry in the context of string theory together with inflation 
is sensitive to 
the values of the light mass, $m_L$, (cf. Eq. (\ref{parametrization})), 
$M_{GUT}$, and the reheating temperature, $T_{RH}$. 
Our results indicate that for rather diverse set of values of these 
parameters \cite{paper} 
we can obtain values for $r$ that are consistent with amplification
via galactic dynamo or collapse of protogalactic clouds.

\end{document}